\documentclass[12pt,a4paper]{article}       

\usepackage{xcolor}
 \usepackage[skins,theorems]{tcolorbox}
\tcbset{highlight math style={enhanced,
  colframe=red,colback=white,arc=0pt,boxrule=1pt}}
  \usepackage[bookmarksopen, bookmarksnumbered, bookmarksopenlevel=2]{hyperref}
 \usepackage[UKenglish]{babel}
 \usepackage[toc,page]{appendix}
 \usepackage{amsmath}
 \usepackage{amssymb}
 \usepackage{graphicx}
 \usepackage{hhline}

 \usepackage[bf]{caption}
\usepackage{cite}
\usepackage[vcentermath]{youngtab}
\usepackage{geometry}
\usepackage{slashed}
\usepackage{stackrel}
\usepackage{mathtools}
\usepackage{cancel} 
\usepackage{multirow}
\usepackage[margin=0pt,font=small,labelfont=normalfont,skip=22pt]{subcaption}

\usepackage{empheq}
\usepackage{arydshln}

\newcommand{\bqa}{\begin{eqnarray}}
\newcommand{\eqa}{\end{eqnarray}}

\newenvironment{eqn*}{\begin{equation*}\begin{aligned}}{\end{aligned}\end{equation*}\noindent}
\hypersetup{
    pdftitle={},
    pdfauthor={},
    pdfsubject={}
}
\numberwithin{equation}{section}
\numberwithin{table}{section}

\makeatletter

\definecolor{dgreen}{rgb}{0,0.45,0.2}
\definecolor{dblue}{rgb}{0,0.0,0.5}

\newcommand{\be}{\begin{equation}}
\newcommand{\ee}{\end{equation}}
\newcommand{\beq}{\begin{equation}}
\newcommand{\eeq}{\end{equation}}
\newcommand{\ba}{\begin{aligned}}
\newcommand{\ea}{\end{aligned}}

\newcommand{\bea}{\begin{eqnarray}}
\newcommand{\eea}{\end{eqnarray}}

\newcommand\bi{\begin{itemize}}
\newcommand\ei{\end{itemize}}

\def\unit{{1\kern-.65ex {\rm l}}}
\def\1{{1\kern-.65ex {\rm l}}}

\newcount\hour \newcount\minute
\hour=\time \divide \hour by 60
\minute=\time
\def\now{%
\ifnum \hour<13
  \ifnum \hour=0 \advance \hour by 12 \number\hour:\else \number\hour:\fi%
     \ifnum \minute<10 0\fi%
     \number\minute%
\ A.M.%
\else \advance \hour by -12 \number\hour:%
  \ifnum \minute<10 0\fi%
  \number\minute%
  \ P.M.%
\fi%
}

\makeatother

\begin{document}

\begin{titlepage}
\begin{center}
\rightline{\small ZMP-HH/24-19}

\vskip 15 mm

{\large \bf
The Dark Dimension and the Grand Unification of Forces
} 
\vskip 11 mm
Jonathan J. Heckman$^{a,b}$, Cumrun Vafa$^c$, Timo Weigand$^d$ and Fengjun Xu$^e$

\vskip 11 mm

{\it \footnotesize $^a$Department of Physics and Astronomy, University of Pennsylvania, Philadelphia, PA 19104, USA}

{\it \footnotesize $^b$Department of Mathematics, University of Pennsylvania, Philadelphia, PA 19104, USA}

{\it \footnotesize $^c$Jefferson Physical Laboratory, Harvard University, Cambridge, MA 02138, USA}

{\it \footnotesize $^d$ II. Institut f\"ur Theoretische Physik, Universit\"at Hamburg, Luruper Chaussee 149,
22761 Hamburg, Germany}

{\it \footnotesize $^e$ Beijing Institute of Mathematical Sciences and Applications (BIMSA), Beijing, 101408, China}

\end{center}
\vskip 17mm

\begin{abstract}
The dark dimension scenario, predicting one extra mesoscopic dimension in the micron range, has emerged by applying various Swampland principles to the dark energy.  In this note we find that realizing the grand unification of gauge forces is highly constraining in this context. Without actually constructing any GUT models, we argue that the mere assumption of grand unification of forces in this scenario, together with the experimental bounds on massive replicas of the Standard Model gauge bosons, predicts an upper bound for the GUT scale, $M_{GUT}\lesssim 10^{16}\ {\rm GeV}$.  Combined with the experimental bound on the proton lifetime, this predicts that the $X$ gauge boson mediating proton decay is a 5d solitonic string of Planckian tension stretched across a length scale $L\sim ({\rm 1-10\ TeV})^{-1}$ ending on gauge branes of the same diameter $\sim L$.  This leads to a mass of $M_X\sim 10^{15}-10^{16}\ {\rm GeV}$. 
 In particular assuming grand unification in the dark dimension scenario results in a tower of Kaluza-Klein excitations of Standard Model gauge bosons on the gauge branes in the 1-10 TeV range.
 This suggests that the diameter/separation $L$ of the gauge branes correlates with both the weak scale $\sim 1/L$ near a TeV {\it and} the GUT scale $\sim M_5^2 L$ at $10^{16}\ {\rm GeV}$.

\end{abstract}

\vfill
\end{titlepage}
\section{Introduction}

The dark dimension scenario \cite{Montero:2022prj} has emerged as a result of combining Swampland principles in the context of dark energy with observational constraints, leading to a specific corner of the quantum gravity landscape \cite{Gonzalo:2022jac,Law-Smith:2023czn,Obied:2023clp,Gendler:2024gdo,Anchordoqui:2022ejw,Anchordoqui:2022txe,Blumenhagen:2022zzw,Anchordoqui:2022tgp,
Anchordoqui:2022svl,Anchordoqui:2023oqm,Noble:2023mfw,Anchordoqui:2023wkm,Anchordoqui:2023tln,Cui:2023wzo,Anchordoqui:2023etp,Anchordoqui:2024akj,Schwarz:2024tet,Anchordoqui:2024gfa,Casas:2024clw}. In particular it predicts one larger extra dimension
in the micrometer range, while other possible extra dimensions are much smaller.
 This micron scale extra dimension is called the dark dimension.  For a review, see \cite{Vafa:2024fpx} (and \cite{Antoniadis:2024sfa} for a different perspective).  
In this note we evaluate the prospects for accommodating the idea of grand unification of forces, as realized in Grand Unified Theories (GUTs), in the context of a dark dimension.
We find that in this scenario, using 
general principles of quantum gravity and experimental observations (in particular bounds on the proton lifetime and on the mass of excited cousins of Standard Model (SM) gauge bosons) 
uniquely predicts a bound on the mass of the GUT gauge boson, denoted by $X$ (as a generic prediction of GUT models), of the order  $M_X\sim 10^{15}-10^{16}\ {\rm GeV}$, and the existence of a Kaluza-Klein (KK) tower of SM gauge bosons at the mass scale of $1\sim 10 \ \rm{TeV}$.  Moreover we learn that the $X$ gauge boson in the 5d effective field theory (EFT) including the dark dimension is necessarily represented by a stretched solitonic string.

The aim of this paper, just as in the original work on the dark dimension scenario \cite{Montero:2022prj}, is {\it not} to actually construct any specific models, but only to exploit general quantum gravity constraints and known experimental bounds to nevertheless arrive at concrete predictions.  The reason for this point of view is twofold: First, despite continuous progress (see, e.g., the reviews \cite{Ibanez:2012zz,Marchesano:2024gul} and references therein), we do not yet have all the required technology in string theory constructions to 
 combine all the ingredients that we need at the same time and in a fully controlled way.  For example we currently have no fully reliable model with complete moduli stabilization and a hierarchically small positive cosmological constant.  Second, even if we found a model which satisfies all constraints, we would not know whether that is the model of our universe, because we do not know how many models exist which satisfy our constraints.  Therefore any predictions based on specific models, rather than general principles, would be subject to this criticism.  Instead if we only impose constraints that  - to the best of our current understanding - any model would have to satisfy, we can make robust predictions.  This is the path we wish to continue in this paper, by applying the  
  lessons we have learned from quantum gravity and Swampland principles to the issue of grand unification of forces, without any specific models in mind. 

The organization of this 
 note is as follows:  In section \ref{sec:review} we briefly review the relevant aspects of the dark dimension scenario.  In section \ref{sec:GUT} we explain how known experimental constraints, combined with quantum gravity constraints and the   idea of grand unification of forces, lead to a prediction of the $X$-boson mass and of the diameter of the Standard Model (SM) brane supporting gauge fields.  In section \ref{sec:Exp} we conclude by discussing some of the experimental consequences of this picture.

\section{Review of the Dark Dimension Scenario}
\label{sec:review}

  The dark dimension scenario is motivated by the idea that the dark energy is a hierarchically small parameter, and any small parameters in a quantum theory of gravity can occur only if there are towers of light states \cite{Ooguri:2006in,Lust:2019zwm} arising from some dimensions becoming large or a critical string becoming light according to the Emergent String Conjecture\cite{Lee:2019wij}.  
  It was shown in \cite{Montero:2022prj} that combined with other observations this is possible in our universe only if one of the extra dimensions is hierarchically larger than the remaining ones and its size lies in the range 
  \begin{equation}
  L_5\sim \Lambda^{-\frac{1}{4}}\sim 1\  \mu m \,,
  \end{equation}
  where $\Lambda$ is the dark energy. Here and in the subsequent equations we omit $O(1)$ numerical factors and only work at the level of orders of magnitude.
 
This leads, in particular, to the prediction of the 5-dimensional Planck scale $M_5$:
\begin{equation}
M_5^3 L_5\sim M_p^2 \quad \Longrightarrow \quad M_5\sim (10^9-10^{10})\ {\rm GeV} \, \sim \Lambda^{\frac{1}{12}} \textrm{ (in Planck units)} \,. 
\end{equation}
In the dark dimension scenario, the SM states should be localized in the 5-th dimension.  Otherwise we would get a KK tower of SM  particles separated by the (meV-eV) mass scale, which is incompatible with experimental facts. Let us denote by $L_{SM}$ the localization length of the SM brane in the 5-th dimension (which can be viewed as a potentially higher-dimensional brane projected to the 5-th dimension). This can be further refined to the diameter, $L_g$, of the brane where the gauge fields live, and the diameter, $L_m$, of the 
 region where charged matter is localized.  In a string theory setup we expect $L_m\leq L_{g}$, as matter arises typically at the intersection locus of gauge branes.  Experiments bound the mass of the KK tower of gauge bosons with quantum numbers of the SM gauge bosons to be bigger than a few TeV \cite{ATLAS:2017fih, CMS:2021ctt,ParticleDataGroup:2022pth}. This  leads to the bound
\begin{equation}
L_{g}\lesssim ({\rm 1-10\ TeV})^{-1} \,.
\end{equation}
For the matter fields, ignoring the Yukawa interactions we expect an approximate GIM symmetry \cite{Glashow:1970gm} of $SU(3)$ flavor symmetry.  The lack of flavor changing neutral currents (FCNC) implies that this  should be a good symmetry up to an energy scale of at least $M\sim 100\ {\rm TeV}$ (see e.g., \cite{Maiani:2013fpa}). In the context of matter fields localized on branes the breaking scale of the GIM symmetry is given by the scale of the brane diameter.  To see this note that the generations of matter arise in these setups by solving for the zero modes of the Dirac operator (as in \cite{Donagi:2008ca,Beasley:2008dc,Beasley:2008kw}).  To leading order the different zero modes are symmetrical, giving rise to an approximate GIM flavor symmetry.  This symmetry is broken at a scale which can probe the distinct zero mode wave functions of matter fields, which corresponds to the scale of the KK modes of the matter brane.  This implies
\begin{equation}
L_m\lesssim ({\rm 100\ TeV})^{-1}.
\end{equation}
In this scenario the KK tower of graviton excitations in the dark dimension is a promising candidate for dark matter, as has been explored in \cite{Gonzalo:2022jac,Law-Smith:2023czn,Obied:2023clp}. Other scenarios for dark matter in this context have also been investigated \cite{Anchordoqui:2022txe,Anchordoqui:2024akj,Anchordoqui:2023tln}.

\section{Grand Unification and the Dark Dimension}
\label{sec:GUT}

One question that has not yet been addressed in the dark dimension scenario is how  - if at all - the idea of grand unification of forces can be accommodated.  Of course we should first ask whether or not grand unification of forces is necessary for a consistent realization of the SM in quantum gravity. Indeed there is no proof that this is the case, and there exist models in string theory with a SM-like spectrum which at least superficially do not seem to unify at higher energies.  Nevertheless the 
 elegance with which the SM fields fit into simple representations of unified gauge groups,
for example in Georgi-Glashow $SU(5)$ \cite{Georgi:1974sy} or in $SO(10)$ GUTs \cite{Fritzsch:1974nn, Georgi:1975qb},
 can be viewed as strong motivation for the existence of
 an underlying GUT structure.

However, we also know that in the context of string theory, unification of forces may not be a 4-dimensional phenomenon and that unification may be realized only at higher energies where the geometry is higher-dimensional (see e.g.\cite{Candelas:1985en,Witten:1985bz,Acharya:2001gy,Donagi:2008ca,Beasley:2008dc,Beasley:2008kw}).\footnote{Unification of forces with a single large extra dimension has also been considered in \cite{Hall:2001xr, Hall:2001xb, Kim:2002im}. The present scenario is different because the Standard Model degrees of freedom are now quasi-localized in the 
5-th dimension.}  This is indeed the typical situation in string theory.  We can even explain the fact that matter representations fit into GUT representations without the associated GUT actually being realized at any scale (see e.g. \cite{Heckman:2009mn}). With these points in mind we nevertheless look for the possibility of realizing a unification of forces in the dark dimension scenario in a sense as close as possible to what we expect of a `standard' unification.  As part of standard unification schemes we require, by definition, the existence of massive gauge bosons, the remnants of the underlying GUT group, like the $X$-bosons, which can mediate proton decay. For example, in an $SU(5)$ GUT the $X$-bosons are heavy vectors transforming as a $({\bf 3},{\bf 2}) + c.c.$ under $SU(3) \times SU(2)$. We furthermore take the mass of these bosons, $M_X$, as the definition of the GUT scale.

On the face of it we seem to be confronted with a puzzle: At least barring substantial threshold corrections from light states beyond the SM, the GUT scale should arise around $M_{GUT}\sim 10^{16}\ \rm{GeV}$.  However, from the point of view of the dark dimension scenario, this is a bit unexpected because this scale is larger than the 5d Planck scale:
\begin{equation}
M_{GUT}\gg M_5 \,.
\end{equation}
One's immediate reaction may be to hope that due to the KK tower of charged states, the running of gauge couplings might change so much that it brings down the unification scale to $10^{10}\ \rm{GeV}$ or lower.  But even if this is the case, it does not solve the problem:  For such a low unification scale, the mass of the $X$-boson gauge field that mediates proton decay would be $M_X\lesssim 10^{10}\ {\rm GeV}$.  This, however, is in contradiction with the current observational bound on the proton lifetime of $10^{34}$ years \cite{Super-Kamiokande:2020wjk}, which leads to $M_{X}\gtrsim 10^{15}-10^{16} \ \rm{GeV}$ (see e.g. \cite{Ohlsson:2023ddi} and references therein for a recent review).

\begin{figure}[t!]
    \centering
    \includegraphics[width=.85\textwidth]{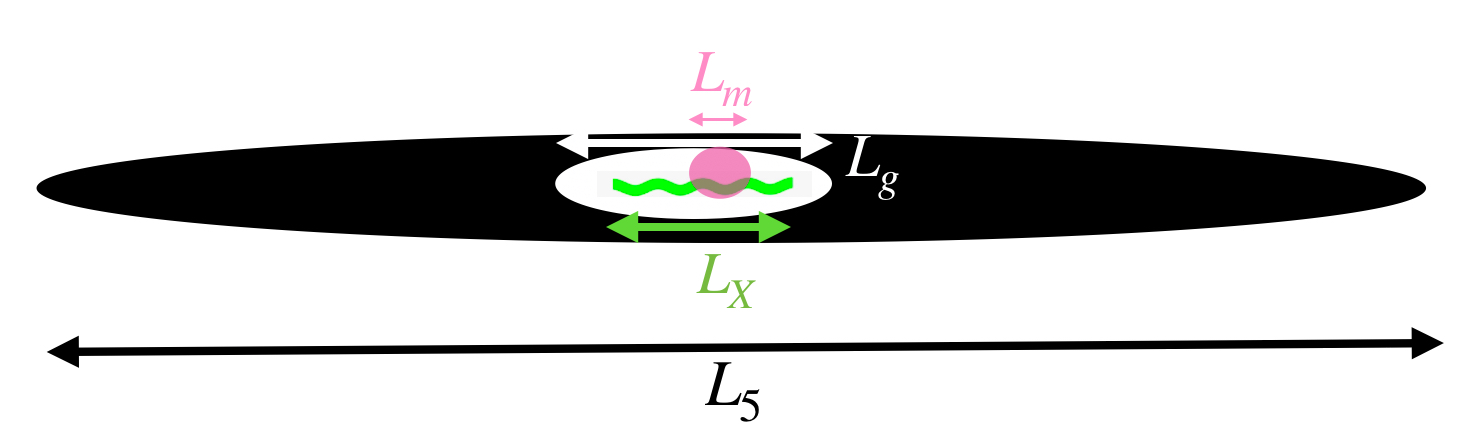}
    \caption{Schematic representation of the dark dimension scenario. The horizontal direction represents the  dark dimension of size $L_5$. The SM gauge fields $SU(3)\times SU(2)\times U(1)$ live in a small region of size $L_g$ in the dark dimension. The pink region of size $L_m$ denotes where the matter fields live. The string represented by the green wavy line stretched along a length of size $L_X$ leads to the $X$-gauge bosons.}
    \label{fig:scenario}
\end{figure}

So it seems we are at an impasse.  Is there any other way to have
 a GUT structure at or above $10^{16}\ \rm{GeV}$?  This would in particular mean that the $X$-boson mass is at $10^{16}\ {\rm GeV}$ or higher.  But no localized object in 5 dimensions can have a mass higher than the Planck scale, without being a black hole!  Hence the only other option is that the $X$-boson is not a point-like object, but an extended object in 5 dimensions (see Figure \ref{fig:scenario}).  In other words, {\it the $X$ gauge boson must be effectively a string in the 5d sense!}  We are assuming that this is a point particle in 4d, as in the GUT unification picture, at least up to the GUT scale.\footnote{Of course we know this cannot be strictly true in the dark dimension scenario because at an energy scale larger than $M_5=10^{10}\ {\rm GeV}$ we are probing the higher-dimensional Planck scale.  However, we can still hope, since we want to assume the grand unification scale at $10^{16} \, {\rm GeV}$, that the 5d Planck scale, which is visible to gravitational terms in the action, is invisible to gauge theoretic interactions in the EFT description, perhaps because the gravitational Planck scale is driven by gauge neutral light states. 
  This is for example what happens in supersymmetric gauge theories as engineered in M-theory on Calabi-Yau threefolds: Near CFT points in their moduli space, a cycle shrinks to zero size in the internal space, 
 and even after becoming much smaller than the Planck scale, it can be unambiguously extrapolated to zero size, as clearly marked by the emergence of massless fields of an effective CFT in the strongly coupled limit.  This is also compatible with the fact that for the EFT in 4d $M_{GUT}<M_{pl}$ and the 5d Planck scale is hard to detect from the 4d perspective \cite{Bedroya:2024uva}.}
 
  This option is of course very natural in string theory.  Indeed, in brane constructions, the gauge bosons are typically obtained by stretched strings between branes.
 We will be assuming this is the usual realization of gauge groups as strings ending on branes. However, we will not be assuming this is a weakly coupled string (and in fact we later argue that it is strongly coupled).

Let $L_X$ denote the length of the string corresponding to the $X$-gauge boson stretched along the dark dimension.  Let the tension of the string be $T$.
Then we have
\begin{equation} \label{eq:MX-1}
M_X\sim T\cdot L_X \,.
\end{equation}
However, for the string itself not to be a black string its tension must be less than 1 in 5d Planck units.  In other words:
\begin{equation}
T\lesssim M_5^2\lesssim 10^{20} \ \rm{GeV}^2 \,,
\end{equation}
where we used the upper range of the estimate for $M_5$ to arrive at an upper bound on $T$.  On the other hand, since the $X$-boson is stretched along the gauge branes it cannot be stretched more than the diameter of the gauge brane.  More precisely, taking for definiteness the example of an $SU(5)$ GUT, the 
 string realizing the $({\bf 3, \bf 2}) + c.c.$
$X$-boson gauge field ends on the $SU(3)$ and $SU(2)$ branes and these two branes must meet also at the matter curves to accommodate the SM matter multiplets.  This implies that the diameter of the total $SU(3)\times SU(2)$ system, $L_g$, should be larger than the string length $L_X$,
\begin{equation}
L_X\lesssim L_g \lesssim (10^4 \ \rm{GeV})^{-1} \,,
\end{equation}
where again we used the upper range of the experimental bounds to deduce an upper bound for $L_X$.
Putting this back into (\ref{eq:MX-1}), 
 we find
\begin{equation}
M_X=T L_X\lesssim {M_5^2}L_g \lesssim \frac{10^{20}\ \rm{GeV}^2}{10^4\ \rm{GeV}}=10^{16} \ \rm{GeV} \,.
\end{equation}
Given the lower bound on the mass of the $X$-boson due to proton decay, we find that the mass of the $X$-boson must lie in the range
\begin{equation}
M_X\sim 10^{15}-10^{16}\ {\rm GeV} \,,
\end{equation}
and that the above inequalities for $T$, $L_X$ and $L_g$ must be equalities:
\begin{eqnarray}
T   &\sim& (10^9- 10^{10} \ \rm{GeV})^2 \,, \\
L_g & \sim& L_X\sim (1-10\ \rm{TeV})^{-1} \,.
\end{eqnarray}
It is quite remarkable that experimental constraints combined with quantum gravity consistency requirements (connecting dark energy and the dark dimension as well as constraints based on black hole physics) have landed us on predictions of various interesting scales in a narrow range, and that this narrow range is close to experimental bounds. It is also natural that we have found that the diameter of the gauge branes, $L_g$, is of the same order as the separation between them, $L_X$.

 Moreover the estimate of $L_g\sim (1-10\ {\rm TeV})^{-1}$ for the scale for the gauge brane diameter was derived without any reference to the weak scale and only by the experimental bound on the mass of the KK tower of the SM gauge bosons.
 It is natural to ask whether the size of the gauge brane is correlated with the weak scale.
 In particular if the Higgs field lives on the gauge brane, then it is natural that the scales associated with the Higgs potential and the weak scale are related to $M_{KK}=1/L_g \sim (1-10)\ \rm{TeV}$.

Indeed this is reminiscent of the gauge-Higgs unification setup \cite{Manton:1979kb,Hosotani:1983xw,Arkani-Hamed:2001nha,Haba:2004bh} where the Higgs field arises from components of a gauge field in higher dimensions, as 
 is the case in string theory.
In that context we have a higher-dimensional gauge theory where the Higgs mass is zero at leading order due to gauge invariance, and a potential is generated due to interactions.  It is easy to see, using the fact that the holonomy is $\theta \sim gH/M_{KK}$, that we generically generate a potential for the Higgs field which has the structure\footnote{Indeed if the holonomy of the higher-dimensional theory is coupled to fermions living on the gauge brane, we expect the mass of the fermions to increase when the holonomy is turned on. This reduces the contribution of the fermions to the vacuum energy and thus results in a decrease in vacuum energy in moving away from the origin of the Higgs field, as is expected for the tachyonic mode of the Higgs field at the origin.}
\begin{equation}
V(H)=M_{KK}^4 f(\theta)= M_{KK}^4 f\left(\frac{g H}{M_{KK}}\right) \,,
\end{equation}
where $g^2\sim 10^{-2}$ is the gauge coupling on the brane.  This leads to the Higgs mass scale 
\begin{equation}
M_H \sim g M_{KK}\sim (10^2-10^3)\ \rm{GeV} \,, 
\end{equation}
which lies in the expected range!

\begin{figure}
\centering
\includegraphics[scale=0.28]{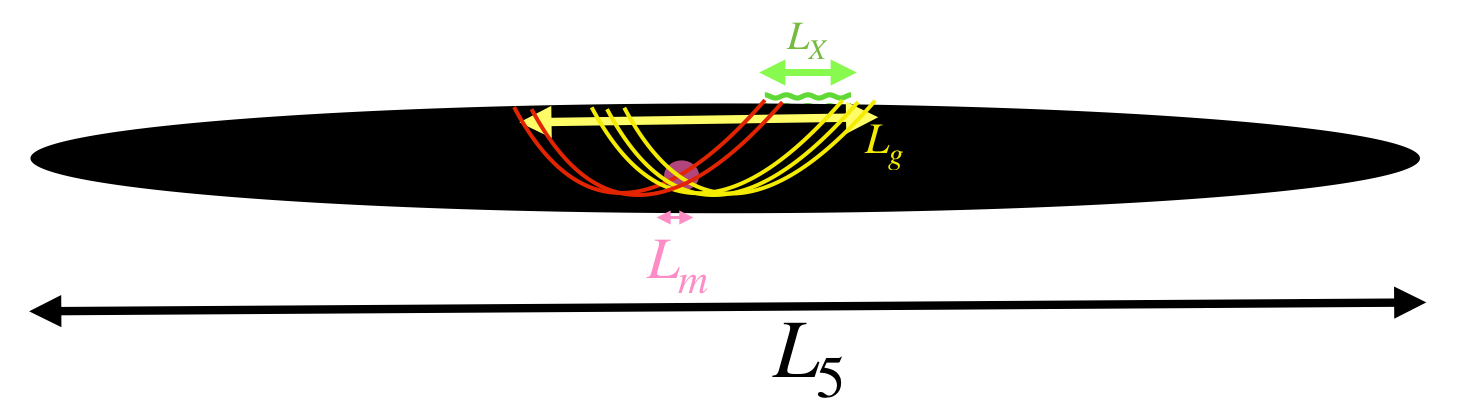}
\caption{A cartoon for a would-be stringy realization where the SM gauge fields $SU(3)\times SU(2)\times U(1)$ live on the branes supported over a small region of size $L_g$ in the dark dimension. They intersect in a region of size $L_m$ where the SM matter fields are localized.  The unification scale is represented by the size of the $X$-boson stretched between the branes.  Such a stringy realization would be at strong coupling, so neither the branes nor their intersection region can be viewed as sharply defined, and should be taken only as a heuristic picture, which could also be hiding additional dimensions.}
\label{fig:GUTbrane}
\end{figure}

One may wonder whether the effective string describing the $X$-gauge boson is a weakly coupled string in stringy realizations of this scenario (see e.g. \cite{Blumenhagen:2000wh,Cvetic:2001nr} for orientifold realizations of GUTs and Fig. \ref{fig:GUTbrane} for an illustration). 
This cannot be the case:  We first have to address how the gauge brane of such a large size can lead to the SM gauge couplings in the right range.   We expect that the 4d gauge coupling can be read off from the effective 5d projection of the gauge brane, leading to gauge kinetic terms in four dimensions of the form
\begin{equation}
c\int F^2 (M_5L_g)\sim c\times 10^6\int F^2 \,, 
\end{equation}
where $c$ represents some prefactor.
This would require $c\sim 10^{-4}$ to arrive at the desired value for $g^2 \sim 10^{-2}$. Since we have identified the tension of the string with $M_5$, 
the prefactor $c$ in such a context would be identified with the inverse of the string coupling $1/\lambda_s$, and so we would be deducing that $\lambda_s \sim 10^4$.  In other words we are in the strong coupling limit of this string.  This suggests we should go to a dual description, which could be either a lighter string or another dimension opening up \cite{Lee:2019wij}. The simpler option is to assume a larger extra dimension.  An example of how such a string can emerge is in terms of M-theory, with a large 11-th circle, where an M2-brane wraps that circle to give an effective string.   We just assumed that the gauge brane would be 5-dimensional.  But it is not hard to see that if we want to embed this scenario in M-theory, it would have to have extra dimensions beyond the ones visible to the 5d observer.  

More generally, how precisely this 5-dimensional solitonic effective string arises from a higher-dimensional object compactified along some of the smaller, extra dimensions is an interesting question
for further and more detailed model building for which we currently lack the full technology.
 As stressed in the Introduction, here we are only trying to constrain what are the allowed models and to check that we are not finding something which is ruled out by some Swampland principles.  It appears that this scenario can in principle be viable and the effective string (from the point of view of the 5-th dimension) giving rise to the $X$-gauge boson is a solitonic string.

In the next section we discuss potential experimental consequences.

\section{Experimental Consequences}
\label{sec:Exp}

It is quite surprising that the scale $M_X\sim 10^{15}-10^{16}\ {\rm GeV}$ for the mass of the $X$-boson  was predicted here not by imposing the correct unification scale for the running couplings, but rather from known experimental bounds combined with quantum gravity consistency arguments.
 What we have shown here is that in the dark dimension scenario, the minimal desert scenario for GUTs, where no new physics occurs in the gauge sector until the GUT scale, is not realized.  In particular, assuming the existence of an $X$-gauge boson led us to the prediction of towers of charged states in the TeV range.  
   In this context if indeed the $X$-gauge boson mass is where the gauge couplings unify, a further explanation is required if or why the new charged states do not dramatically modify the unification scale.

The fact that the scale $L_g$ is close to the scale being probed at LHC is quite interesting.  In particular the scenario considered in this note predicts a KK tower for the SM gauge bosons in the small TeV range, which should be observable in the near future collider experiments.  Indeed it is natural to ask if KK towers involving massive cousins of $W$- and $Z$-bosons, gluons and photons have already impacted observations at the colliders.  For example could they be related to the $W$-boson mass anomaly \cite{CDF:2022hxs}?  Indeed some works hint in this direction
(see e.g. \cite{Strumia:2022qkt,Endo:2022kiw}).

 We can also see that dark energy, GUT physics and the weak scale all come together!  Indeed the fact that unification of the couplings occurs near the Planck scale can be seen by noting that in Planck units, $M_5\sim \Lambda^{\frac{1}{12}}$ and $\Lambda_W\sim M_{KK}\sim \Lambda^{\frac{1}{6}}$ (which is valid up to order one factors) leading to $M_X\sim \frac{M_5^2}{\Lambda_W}\sim 1$ in Planck units (up to corrections in the above equalities bringing it down by about $10^{-4}$).
In other words the 5d Planck scale is approximately the geometric mean of the GUT scale and the KK scale of the gauge brane (echoing the fact that the weak scale is the geometric mean of the 5d Planck scale and the KK scale of the dark dimension $\Lambda^{\frac{1}{4}}$). Moreover, 
the ${\rm TeV}^{-1}$ scale separation and size of the gauge branes
not only set
the GUT scale at $10^{16}$ GeV, but may also underlie the Higgs VEV being close to the TeV scale.

Note that our arguments have nothing to say about the diameter of the matter locus, $L_m$.
Indeed it is very natural that $L_m\ll L_g\sim (1-10 \ \rm{TeV})^{-1}$, as the charged matter is localized in the region of intersection of the gauge branes (see Fig. \ref{fig:GUTbrane}).  Thus the FCNC suppression is natural in this setup.

The thickness of the gauge boson brane will have some impact on other phenomenological aspects of the dark dimension.  
 For example, the bound $f_a\lesssim {M_5}\sim 10^{10}\ {\rm GeV}$ for the axion decay constant in \cite{Gendler:2024gdo}  was found under the assumption that the thickness of the SM brane is Planckian in 5d; here this gets modified by a factor of $\sqrt{M_5L_g}\sim 10^3$
and becomes $f_a\lesssim {M_5}\times 10^3 \sim 10^{13}\ {\rm GeV}$,
which sits comfortably in the allowed experimental region.
It would also be interesting to study the impact of this scenario on the cosmology of the dark dimension scenario \cite{Gonzalo:2022jac}. 

Finally it would be interesting to try and find pathways within string theory where we can realize this scenario, perhaps along the lines initiated in \cite{Schwarz:2024tet}.  Of course one should also keep in mind, as in many string theory realizations, that the GUT-like representations of matter can emerge without any grand unification of forces.  We are open to either possibility.  This paper shows that in the dark dimension scenario, we can in principle accommodate some aspects of GUT predictions such as the existence of the $X$-gauge bosons, but at the very least we have to abandon the desert scenario for GUTs and expect interesting towers of KK modes for SM gauge bosons in the TeV range.

\subsubsection*{Acknowledgments} 
We would like to thank Alek Bedroya, Howard Georgi, Georges Obied, Lisa Randall and Matt Reece for interesting discussions.  We also thank the SCGP for hosting the 2024 Physics summer workshop, where this work was initiated. The work of J.J.H is supported by DOE (HEP) Award DE-SC0013528 and BSF grant 2022100. The work of C.V. is supported in part by a grant from the Simons Foundation (602883,CV) and the DellaPietra Foundation. The work of T.W. is supported in part by Deutsche Forschungsgemeinschaft under Germany's Excellence Strategy EXC 2121  Quantum Universe 390833306, by Deutsche Forschungsgemeinschaft through a German-Israeli Project Cooperation (DIP) grant ``Holography and the Swampland" and by Deutsche Forschungsgemeinschaft through the Collaborative Research Center SFB1624 ``Higher Structures, Moduli Spaces, and Integrability’'. The work of F.X. is supported by the Young Scientists Fund of NSFC under the grant No.12205159.

\newpage

\bibliographystyle{jhep}

\providecommand{\href}[2]{#2}\begingroup\raggedright\endgroup

\end{document}